\documentclass{aa}  

\usepackage{graphicx}
\usepackage{txfonts}
\usepackage{natbib}

\begin{document}

   \title{Rings and filaments: The remarkable detached CO shell of U Antliae\thanks{This paper makes use of the following ALMA data: ADS/JAO.ALMA2015.1.00007.S. ALMA is a partnership of ESO (representing its member states), NSF (USA) and NINS (Japan), together with NRC (Canada), NSC and ASIAA (Taiwan), and KASI (Republic of Korea), in cooperation with the Republic of Chile. The Joint ALMA Observatory is operated by ESO, AUI/NRAO and NAOJ.}$^,$ \thanks{The reduced ALMA FITS data cubes used in this paper are available in electronic form at the CDS via anonymous ftp to cdsarc.u-strasbg.fr (130.79.128.5) or via http://cdsweb.u-strasbg.fr/cgi-bin/qcat?J/A+A/t}}

   \author{F. Kerschbaum\inst{1}, M. Maercker\inst{2}, M. Brunner\inst{1}, M. Lindqvist\inst{2}, H. Olofsson\inst{2}, M. Mecina\inst{1},  
E. De Beck\inst{2}, M.A.T. Groenewegen\inst{3}, E. Lagadec\inst{4}, S. Mohamed\inst{5,6},
C. Paladini\inst{7}, S. Ramstedt\inst{8}, W. H. T. Vlemmings\inst{2}, M. Wittkowski\inst{9}
          }
          
\offprints{Franz Kerschbaum, University of Vienna, Austria, \email{franz.kerschbaum@univie.ac.at}}

   \institute{
Department of Astrophysics, University of Vienna, Türkenschanzstr. 17, 1180 Vienna, Austria
\and
Department of Earth and Space Sciences, Chalmers University of Technology, Onsala Space Observatory, 43992 Onsala, Sweden
\and
Koninklijke Sterrenwacht van Belgi\"{e}, Ringlaan 3, B-1180 Brussels, Belgium
\and
Laboratoire Lagrange, Universit\'{e} C\^{o}te d’Azur, Observatoire de la C\^{o}te d’Azur, CNRS, Bd de l’Observatoire, CS 34229, F-06304, Nice Cedex 4, France
\and
South African Astronomical Observatory, P.O. Box 9, 7935 Observatory, South Africa
\and
Astronomy Department, University of Cape Town, University of Cape Town, 7701, Rondebosch, South Africa National Institute for Theoretical Physics, Private Bag X1, Matieland, 7602, South Africa
\and
Institut d’Astronomie et d’Astrophysique, Universit\'{e} Libre de Bruxelles, Campus Plaine C.P. 226, Boulevard du Triomphe, B-1050 Bruxelles, Belgium
\and
Department of Physics and Astronomy, Uppsala University, Box 516, 75120 Uppsala, Sweden
\and
European Southern Observatory, Karl-Schwarzschild-Straße 2, 85748 Garching, Germany
}

   \date{Received 21 February, 2017; accepted 4 August, 2017}

 
  \abstract
   {}
   {Our goal is to characterize the intermediate age, detached shell carbon star U Antliae morphologically and physically in order to study the mass-loss evolution after a possible thermal pulse.}
   {High spatial resolution ALMA observations of unprecedented quality in thermal CO lines allow us to derive first critical spatial and temporal scales and constrain modeling efforts to estimate mass-loss rates for both the present day as well as the ejection period of the detached shell.}
   {The detached shell is remarkably thin, overall spherically symmetric, and shows a barely resolved filamentary substructure possibly caused by instabilities in the interaction zone of winds with different outflow velocities. The expansion age of the detached shell is of the order of 2700 years and its overall width indicates a high expansion-velocity and high mass-loss period of only a few hundred years at an average mass-loss rate of $\approx$\,10$^{-5}$\,$M_\odot$\,yr$^{-1}$. The post-high-mass-loss-rate-epoch evolution of U~Ant  shows  a significant decline to a substantially lower gas expansion velocity and a mass-loss rate amounting to 4$\times$10$^{-8}$\,$M_\odot$\,yr$^{-1}$ , at present  being consistent with evolutionary changes as predicted for the period between thermal pulses.
   
   }
{}


\titlerunning{The remarkable detached shell of U Antliae}
\authorrunning{F. Kerschbaum et al.}
   \maketitle
%

\section{Introduction}

Mass loss on and shortly after the asymptotic giant branch (AGB) is a major factor in the 
post-main-sequence evolution of low- and intermediate-mass stars. The photospheres and expanding 
outflows show a rich molecular chemistry, and are the formation sites for microscopic dust 
particles. Mass loss plays a crucial role within the cosmic cycle of matter, providing up to 70\,\%
of the interstellar dust \citep[e.g.][]{Schneider2014}. 
Understanding the chemistry in the atmospheres and circumstellar envelopes (CSEs) of AGB stars, the mass loss, and the mixing of material into the interstellar medium (ISM) is therefore crucial to understanding early star formation and the evolution of galaxies. Studies of AGB stars are also important for understanding the future of our solar system.

Mass loss of 
AGB stars takes the form of slow (typically 5--25 km/s) winds with high mass-loss rates (of up to 10$^{-4}$ M$_\sun$/yr).
Although mass loss has been studied since the 1960s, many basic questions on its temporal evolution, geometry, and dynamics remain unanswered. 

There is growing evidence that mass loss on the AGB is not a continuous process. One cause for variations are He shell flashes (thermal pulses), which occur typically every 10$^4$ to 10$^5$ years \citep{Vassiliadis1993}. 
They are thought to be responsible for the formation of detached shells observed around carbon AGB stars \citep{Olofsson1996}. High spatial resolution observations of these objects have shown thin, clumpy, detached shells, indicative 
of short phases of intense mass loss, colliding with a previous, slower AGB wind 
\citep{Lindqvist1999, Olofsson2000, Schoeier2005, Maercker2012}.

The presence of a binary companion can further complicate outflows from AGB stars \citep[e.g.][and references therein]{Mayer2011,Maercker2012,Ramstedt2014}, and high spatial resolution observations of gas and dust (in mm-CO, far-IR and visual scattered light) are 
needed to determine morphology and mass-loss-rate evolution.
The same holds for the clumpiness of the detached shells and the relative location of dust and gas 
\citep{Kerschbaum2010,Olofsson2010,Maercker2010,Maercker2016}, since they potentially 
hold crucial information on the mass-loss mechanism and the physical processes predominating in the wind.
    
The carbon star U Antliae provides a special opportunity to study these questions in detail. Known already for its far-IR excess in the late 1980s and imaged using the Infrared Astronomical Satellite (IRAS) \citep{Waters1994,Izumiura1995}, 
it was evident that this object (260\,pc from {\it Hipparcos}) had a spatially extended detached dust shell with a 
radius of about 50\arcsec. Single-dish mapping in the $^{12}$CO(1--0) and (2--1) lines revealed a detached shell of radius $\sim$43\arcsec{}\ \citep{Olofsson1996}. 
Observations of scattered stellar light around U Ant \citep{Gonzalez2001} revealed a complex shell 
structure. The authors identified four shells at $\sim$25\arcsec, 37\arcsec, 43\arcsec, and 46\arcsec, referred to as shell 1, 2, 3, and 4 below. This  structure was confirmed by studying scattered polarized light \citep{Gonzalez2003, Maercker2010}. The mass of the outermost shell 4 (after \citet{Maercker2010} more at 50\arcsec{}) is dominated by dust, while shell 3 at 43\arcsec{}\ mainly consists of gas. Radiative transfer models of single-dish CO data are consistent with a thin CO shell at 43\arcsec{}\ \citep{Schoeier2005}. 

The peak intensity of the thin (<3\arcsec) dust shell seen using 
Herschel-PACS at 70\,$\mu$m \citep{Kerschbaum2010} coincides with shell 3, which is attributed to material expelled $\sim$3000\,years ago. This age places U~Ant just between other detached-shell objects that were observed by mm-interferometers in the last decade: U~Cam ($\sim$800\,years, 
observed with the Institut de radioastronomie millimétrique Plateau de Bure interferometer (IRAM PdB); \citet{Lindqvist1999}), R~Scl ($\sim$2300\,years, observed with Atacama Large Millimeter/submillimeter Array (ALMA); 
\citet{Maercker2012,Maercker2016}), and the more evolved shell of TT~Cyg ($\sim$7000\,years, observed with IRAM PdB; \citet{Olofsson2000}). 

Observations by \citet{Maercker2010} also provide information on light scattered by Na and K atoms in the shell. Such data sets provided the highest spatial resolutions so far. Nevertheless, a complete picture is hampered by the low spatial resolution of the single dish mm-CO data, and the lack of velocity information in the optical and IR observations.

   \begin{figure*}
   \centering  
\resizebox{\hsize}{!}{\includegraphics[clip, trim = 8.1cm 8cm 4.5cm 8cm, width=1.00\textwidth]{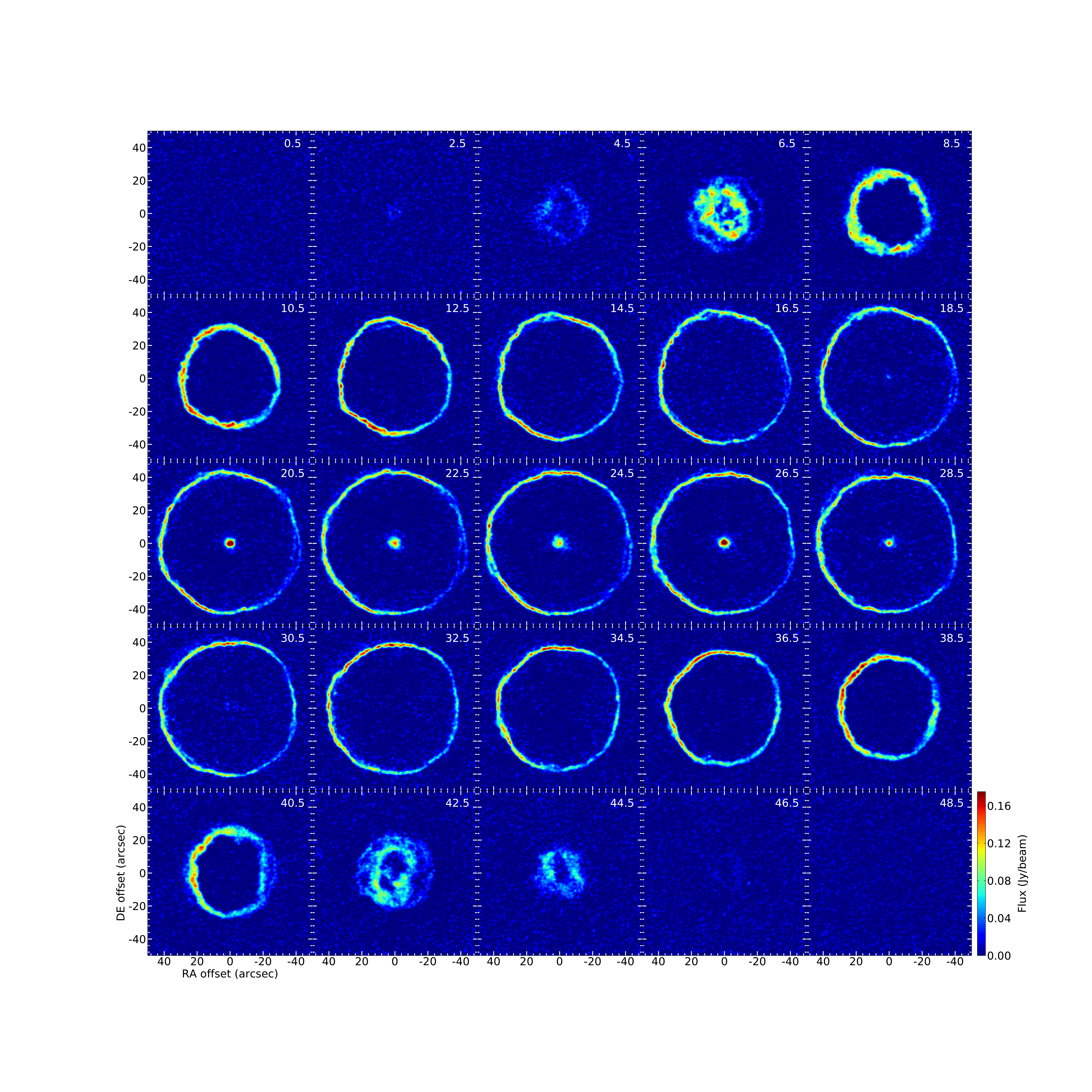}}
   \caption{ALMA (MA+ACA+TP) observations of $^{12}$CO(1--0) towards U Ant. The central LSR velocity in units of km/s of each subpanel is indicated in the respective upper right corners. The velocity resolution is binned to 1\,km/s. The colour bar  to the right gives the flux density scale.}
              \label{F1_B3_overview}%
    \end{figure*}

   \begin{figure*}
   \centering
\resizebox{\hsize}{!}{\includegraphics[clip, trim = 8.1cm 8cm 4.5cm 8cm, width=1.0\textwidth]{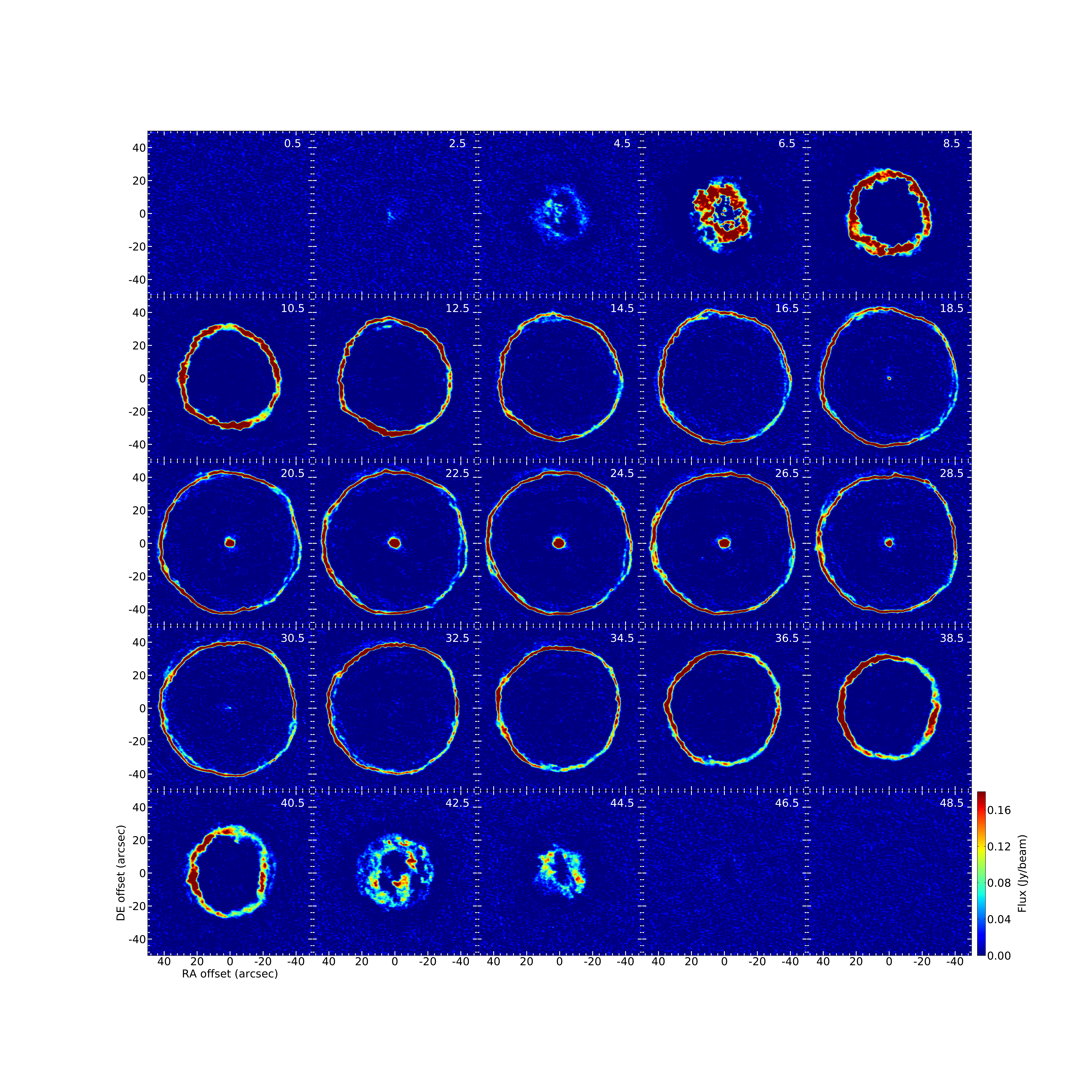}}
   \caption{Same as Fig.~\ref{F1_B3_overview} for the $^{12}$CO(2--1) line.}
              \label{F2_B6_overview}%
              
\resizebox{\hsize}{!}{\includegraphics[clip, trim = 0cm 0cm 0cm 0cm, width=1.0\textwidth]{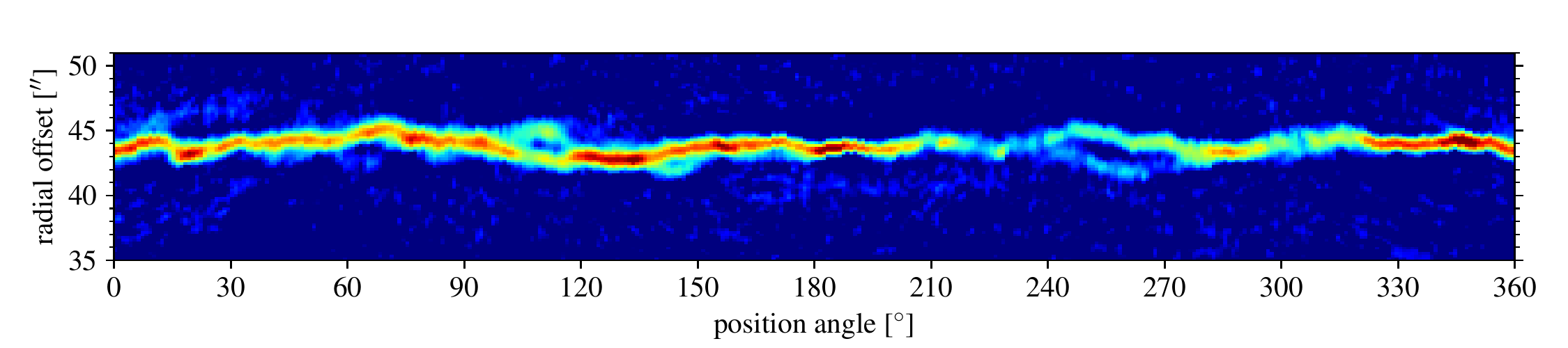}}

   \caption{ALMA (MA+ACA+TP) observations of $^{12}$CO(2--1) towards U Ant spectrally integrated close to the systemic velocity (23.5--25.5 km/s), mapped to polar coordinates, and shown only for radial distances close to the detached shell (compare also Fig.~\ref{F3_B6_PACS_pol} for a similar representation of the whole envelope incl.~dust emission.)}
              \label{F2_B6_polar}%
              
    \end{figure*}

\section{Observations and data reduction}
\label{sec:observations}

ALMA observations hold the potential to change this situation dramatically, delivering high sensitivity observations of the circumstellar environment of U Ant at high angular and spectral resolution.
We observed U\,Ant in ALMA Cycle 3 using an intermediate and a compact 12\,m-array (MA) configuration in band 3 (B3) and band 6 (B6), targeting the $^{12}$CO(J=1--0) and $^{12}$CO(J=2--1) lines. The CSE of U\,Ant was covered by a mosaic of 23 and 85 individual pointings in B3 and B6, respectively, resulting in a map of 110\arcsec$\times$110\arcsec. Additionally, short-baseline observations with the 7\,m-array (ACA) and single-dish measurements with the total power (TP) array to recover all spatial scales were carried out. 
The beam sizes in the reduced maps are 1.82\arcsec $\times$1.63\arcsec {}~(PA=80.5\degr{}) for CO(1--0)
and  1.37\arcsec $\times$1.00\arcsec  {}~(PA=86.6\degr{}) for CO(2--1).
In B3 the main target of the spectral setup was the CO(1--0) line at 115.271\,GHz, using a spectral resolution of 244.019\,kHz, which corresponds to a velocity resolution of 0.635\,km/s. In B6 the main target was the CO(2--1) line at 230.538\,GHz, observed with a resolution of 488.037\,MHz to achieve the same velocity resolution as in B3. 

   \begin{figure*}
   \centering
  \resizebox{\hsize}{!}{\includegraphics[clip, trim=-2cm 0cm -2cm 0cm, width=1.0\textwidth]{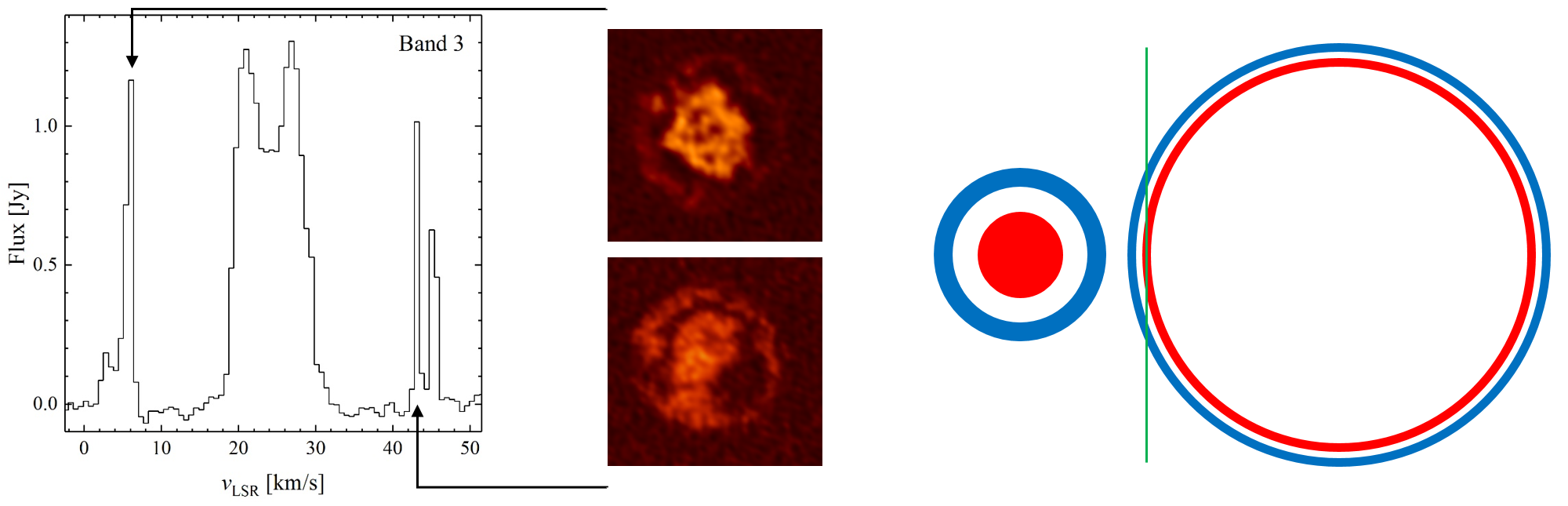}}
   \caption{Left: Integrated B3 $^{12}$CO(1--0) spectrum obtained in a circular, 7\arcsec{}\ wide region centred on U Ant. Middle: Enlarged 0.65\,km/s resolution velocity channel maps at $\pm$18.5\,km/s relative to the systemic velocity. Right: Illustration of the geometry when selecting special velocity channel maps of a double shell structure, first as seen by the observer as in the middle maps and second in a hypothetical side view with the green line indicating the spatial cut corresponding to the chosen velocity channel.}
              \label{F4_B3_cups}%
    \end{figure*}
    
In order to optimize the final data product quality, we re-calibrated and re-imaged the data. The calibration was done with the Common Astronomy Software Application (CASA) pipeline scripts as delivered to the principal investigator and no major adjustments needed to be done.
Flux calibrators for the individual data sets were Mars and Ganymede; phase calibrators were the quasars J1042-4143 and J1107-4449. The calibrated data sets of the MA and ACA were combined in the uv-plane with appropriate weighting. 

The imaging process, which is critical for the final data quality, was carried out with extended and custom python imaging scripts by our team; these scripts were executed in CASA. These scripts make use of an iterative and interactive imaging procedure with the CLEAN algorithm, using Briggs weighting (with robustness = 0.5), multi-scale cleaning, and iterative masking with decreasing thresholds to deconvolve the inter-shell structures of the CSE. 
Compared to the imaging scripts of the delivered data products, our strategy employs more mask iterations with finer thresholds and the CLEAN masks are carefully adapted where necessary.

After the imaging process for the interferometric data, the TP single-dish data were added via feathering.\footnote{Although feathering of single-dish to interferometric data is known to be problematic and in most cases slightly incorrect, up to now no other, more reliable methods to combine single-dish data in the visibility plane are currently under development. The authors are aware of the fact that the resulting images may contain artefacts, but thorough analysis suggests that the TP inclusion via this process does not alter the data significantly.} 
For the final data products presented in Figs.~\ref{F1_B3_overview} and \ref{F2_B6_overview}, velocity binning to 1\,km/s was applied during the imaging to increase the signal-to-noise ratio in the fainter parts of the shell.

\section{Interpretation}

\subsection{Morphology of the detached shell}

A first inspection of the B3 and B6 maps close to the systemic velocity of 24.5\,km/s in Figs.~\ref{F1_B3_overview} and \ref{F2_B6_overview} reveals a remarkably thin, overall spherically symmetric, detached CO shell of about 42.5\arcsec{}\ radius expanding at a velocity of about 19\,km/s. At a distance of 260\,pc this corresponds to an expansion age of about 2700 years and a physical radius of 0.05\,pc. The shell appearance is quite similar in both CO lines. The detached shell centre coincides with the stellar position within the measurement errors. 

On closer inspection, the detached shell shows some fine, filamentary structures  down to the resolution limit of about 1.5\arcsec. In several regions the shell is split into two basically unresolved filaments differing in radius by up to 5\arcsec. This splitting is seen in many of the velocity channels and seems to be a general property of the shell as it is supported by the B6 spectral scan movie in Fig.~\ref{F6_B6_movie} in the appendix. Figure~\ref{F2_B6_polar} makes this even more evident by plotting the CO(J=2--1) emission of the detached shell close to systemic velocity in polar coordinates. While at other velocities the details of the shell change, the overall filamentary structure remains.

The partially split structure is further seen through a comparable phenomenon at the extreme velocity channel maps both in Figs.~\ref{F1_B3_overview} and \ref{F2_B6_overview}, and in the integrated spectrum close to the stellar position in Fig.~\ref{F4_B3_cups}. The latter reveals a split in velocity by about 3\,km/s at both velocity extremes. This split is also responsible for the double ring structure seen in the middle panel of Fig.~\ref{F4_B3_cups}, which shows enlarged 0.65\,km/s resolution velocity channel maps at $\pm$18.5\,km/s relative to the systemic velocity. The illustration on the right side of Fig.~\ref{F4_B3_cups} visualizes the geometrical situation when selecting a velocity channel map at the extreme velocity of the ``inner'' shell filament leading to a spatially separate cut through the ``outer'' shell filament, consistent with the data shown in the middle panel where a filled circle is surrounded by a ring. 

The question is whether we see the same phenomenon in the channel maps and in the spectrum? 
A plausible interpretation is that the splitting into filaments is a result of the supersonic collision of a fast wind resulting from the thermal pulse and a slower normal AGB wind. The gas at the interface heats up and expands outwards resulting in the creation of a reverse shock (moving towards the star) and a forward shock (expanding into the pre-thermal pulse material). Depending on the gas densities and velocities, the thin shell that forms from this supersonic interaction can be subject to dynamical and/or thermal instabilities \citep[e.g.][]{Mohamed2012, Mohamed2013}. The filamentary structure we observe may result from the growth of such instabilities. 

This is further supported by the coincidence that a velocity splitting as observed in the extreme velocity channels shown in Fig.~\ref{F4_B3_cups} and discussed above could also explain the spatial splitting into filaments of up to 5\arcsec{} in the stellar velocity channel image over a timescale comparable to the geometric expansion age of the shell (i.e. 2700 yr). 

\citet{Mohamed2012} consider the collision of a stellar wind emanating from a runaway star with the flow from the ISM. The collision produces the same type of double-shocked structure described above and their ``slow'' models have similar gas dynamical properties to our study. In their Fig.~B.2., the emissivity cross-section map due to rotational transitions of CO demonstrates how the molecular emission in the forward shock differs from that in the reverse shock; the emission is strong in the denser, shocked stellar wind and weaker in the shocked ambient medium. 

The growth of instabilities (e.g. Rayleigh-Taylor instabilities) is also demonstrated and results in mixing between the two layers and the formation of filamentary substructure when projected on the sky (e.g. see their Fig.~12). The 3D appearance of these instabilities is comparable in structure to the filaments we observe.
A definite, physical interpretation of the detached shell fine structure has to await full hydrodynamical simulations foreseen for an upcoming paper. 
 
\begin{figure}
\resizebox{\hsize}{!}{\includegraphics[clip,trim=1.3cm 3cm 1.3cm 3.9cm,width=1.0\textwidth]{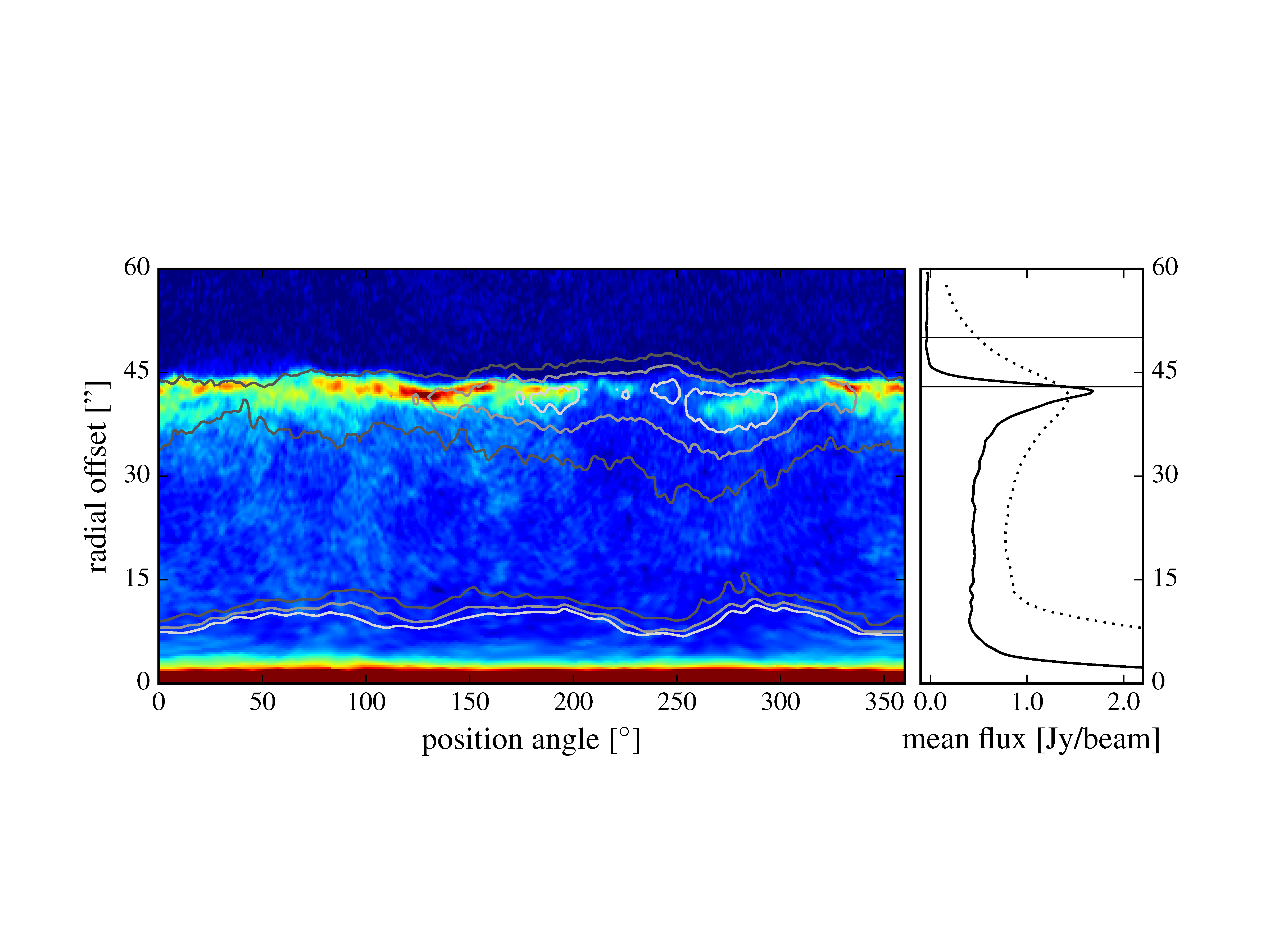}}
  \caption{U Ant $^{12}$CO(2--1) spectrally integrated data mapped to polar coordinates over plotted with PACS 70\,$\mu$m contours.  The  graph to the right depicts azimuthally averaged flux in the ALMA (full) and the Herschel (dashed) data. The two horizontal lines at 43 and 50\arcsec{}\ indicate the radial positions of shells 3 and 4 as discussed in \citet{Maercker2010}.}
  \label{F3_B6_PACS_pol}
\end{figure}

One can also compare our gas shell morphology with that of the dust. Figure~\ref {F3_B6_PACS_pol} shows spectrally integrated (moment 0) $^{12}$CO(2--1) data, mapped onto polar coordinates with 
Herschel PACS 70\,$\mu$m contours \citep{Kerschbaum2010,Cox2012}. Overall, the spatial distributions are quite similar and resemble each other even in several details, indicating a strong coupling between gas and dust. 

Nevertheless, it should be noted that the brightest regions in dust emission at position angles between 160 and 300\degr{}  correspond to an overall fainter region in gas emission -- something that was already evident when comparing the PACS data to earlier single dish CO measurements of U Ant \citep{Olofsson1996}. This partial anti-correlation is surprising since possible photodissociation effects would act the other way round with a positive correlation by the shielding from dust. Moreover, Fig.~\ref {F3_B6_PACS_pol} indicates that the molecular gas shell as observed by ALMA is co-spatial with the gas dominated shell 3 at 43\arcsec{}\ of \citet{Maercker2010} and not their dust shell 4 further out. 

In \citet{Cox2012} a U Ant space velocity of 36\,km/s, with respect to the ISM, at a projected position angle of 285.8\degr{} was derived, as well as a 5.3\arcmin{}\ stand-off-distance of a possible bow shock. In Fig.~\ref {F3_B6_PACS_pol} this position angle corresponds to a region of somewhat smaller detached shell radius. The  very different spatial scales (5.3\arcmin{}\ versus the 43\arcsec{}\  shell radius) nevertheless make a connection unlikely.
One should also  note that we see no signs of shells 1 and 2 in our ALMA CO data. So far this remains a mystery since something is clearly seen in the Na data (but not in the K data) of \citet{Maercker2010}. 

Moreover, all our global spectra are asymmetric, with the blue-shifted side emitting more strongly. The profiles are surprisingly similar to those of R Scl \citep{Maercker2016}. This asymmetry was noted already by \citet{Olofsson1996} for R~Scl, U~Ant, and S~Sct. The star TT~Cyg seems to be an exception. Keeping in mind the still small sample size, however, this may be a pure coincidence.

\subsection{Recent and older mass loss}

\begin{figure}
  \resizebox{\hsize}{!}{\includegraphics[clip, trim=7.1cm 5.5cm 3.6cm 7.6cm, width=1.0\textwidth]{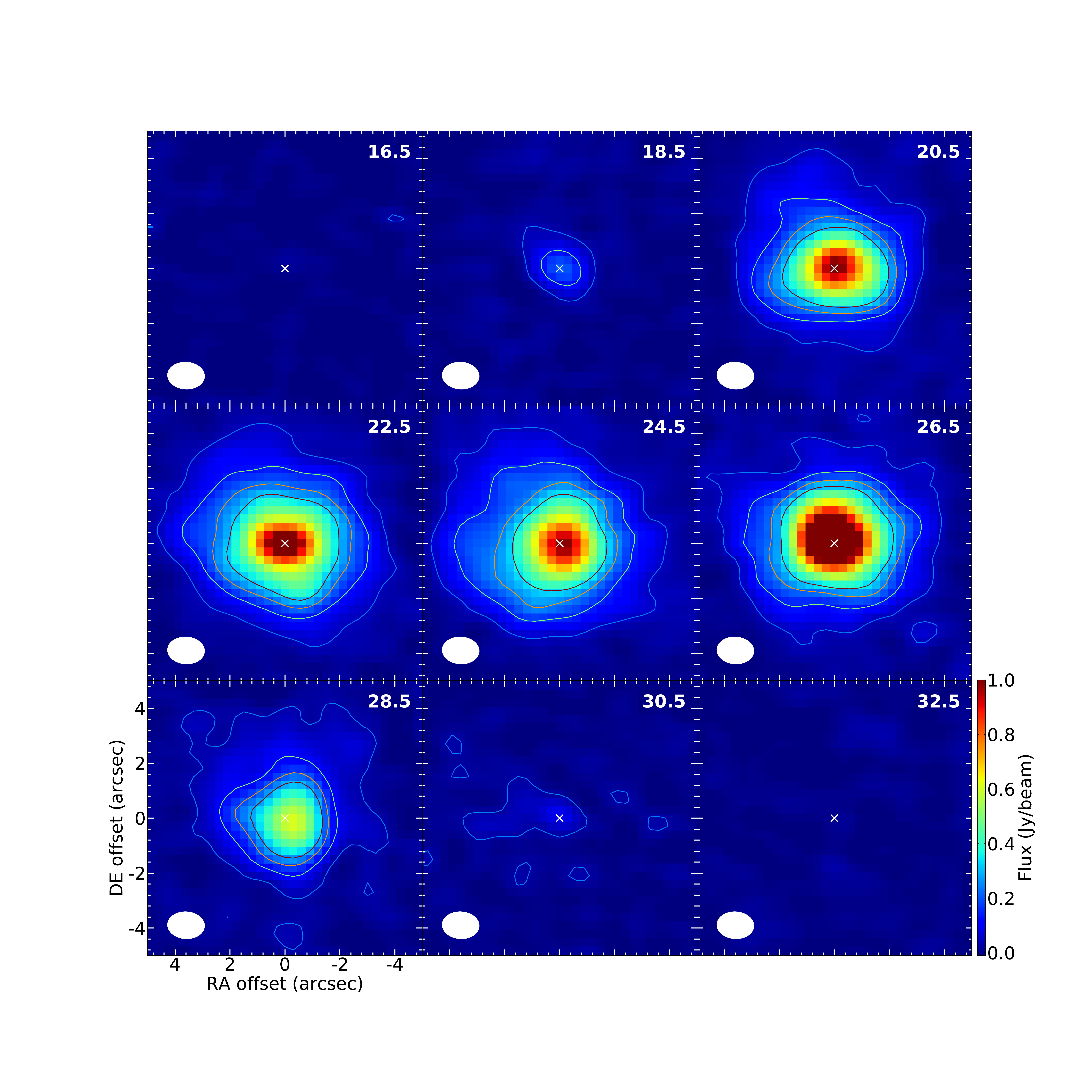}}
  \caption{ALMA observations in the $^{12}$CO(2--1) line of central regions of selected velocity sub panels of Fig.~\ref{F2_B6_overview} (see Fig. 2 for further description)  }
  \label{F2c_B6_centre}
\end{figure}

When zooming in closer to the star, it is evident from our maps (Figs.~\ref{F1_B3_overview}, \ref{F2_B6_overview}, and \ref{F2c_B6_centre}) that the detached shell is not completely hollow. There is new present-day mass loss confined to a region of about 6\arcsec{}\  in radius. There may even be slight deviations from spherical symmetry, unfortunately at the limit of the spatial resolution as indicated by our ALMA beam size in Fig.~\ref{F2c_B6_centre}. 

\begin{figure}
  \resizebox{\hsize}{!}{\includegraphics[clip, trim=3.4cm 4.9cm 1.0cm 4.0cm, width=1.0\textwidth]{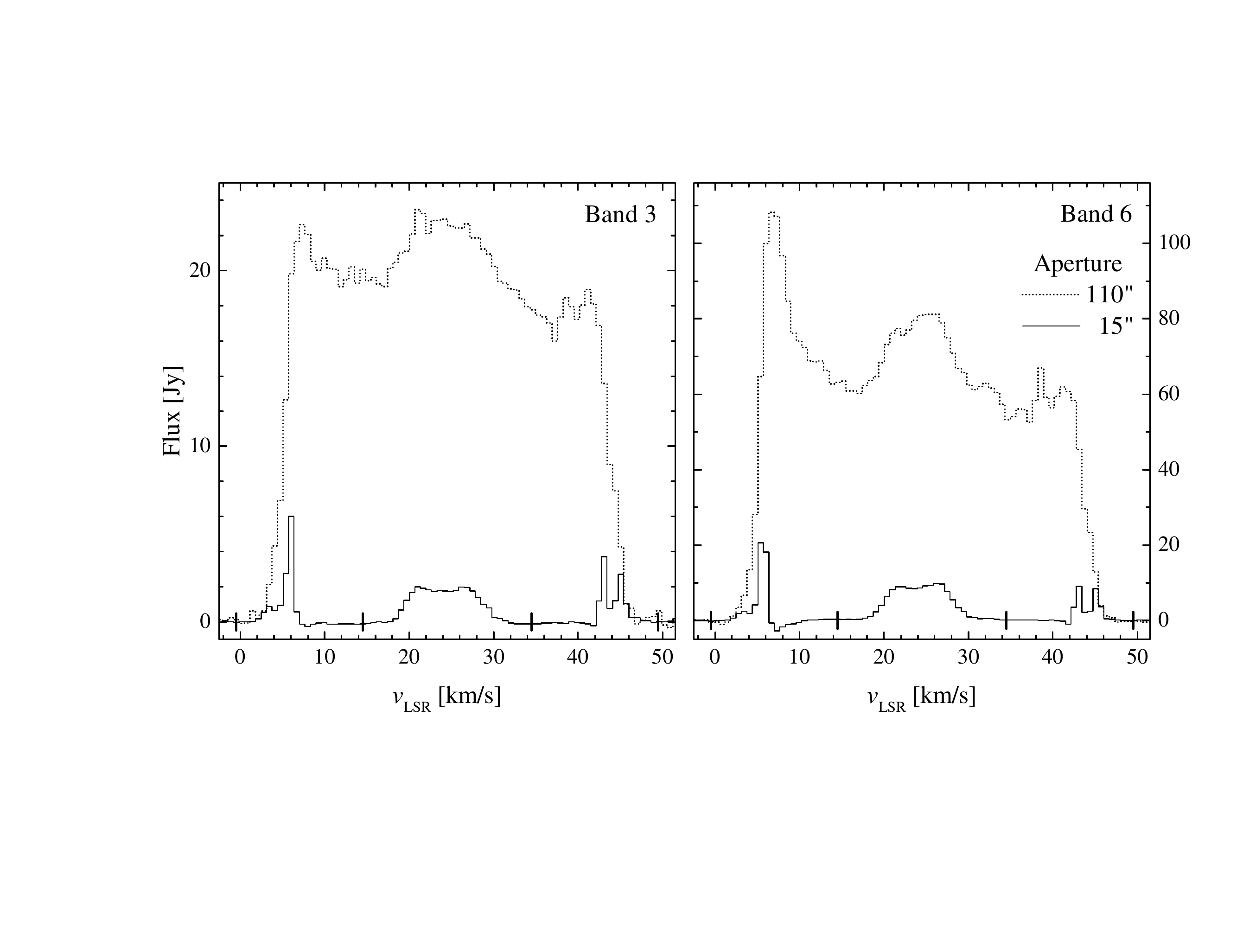}}
  \caption{Spatially integrated $^{12}$CO(1--0) and $^{12}$CO(2--1) spectra using two different aperture diameters fitting the whole emitting region and only the region showing new mass loss close to the star.}
  \label{F5_B36_Spectra}
\end{figure}

As seen in Fig.~\ref{F5_B36_Spectra}, this central region, covered by a 15\arcsec{}\  aperture, shows much smaller expansion velocities of only 4.7 and 4.2\,km/s in B3 and B6, respectively, than the spectra that were produced by using a 110\arcsec{}\ aperture covering the whole emitting region. An expansion age of the present-day mass loss of less than 1600\,yr can be estimated from these numbers.  

This very low present-day expansion velocity is well in line with earlier findings on DR~Ser, V644~Sco \citep{Schoeier2005}, TT~Cyg \citep{Olofsson2000}, and S~Sct \citep{Olofsson1996}, and as expected from the interpulse expansion velocity evolution predicted by \citet{Vassiliadis1993}. The relatively high present-day expansion velocity of U~Cam \citep{Lindqvist1999} could be explained by its very young detached shell (only 700\,yrs as compared to the several thousands of most of the others). The binary R~Scl may also be a special case \citep{Maercker2016}.

\begin{figure*}
  \resizebox{\hsize}{!}{\includegraphics[clip, trim=-17cm 0cm -17cm 0cm, width=1.0\textwidth]{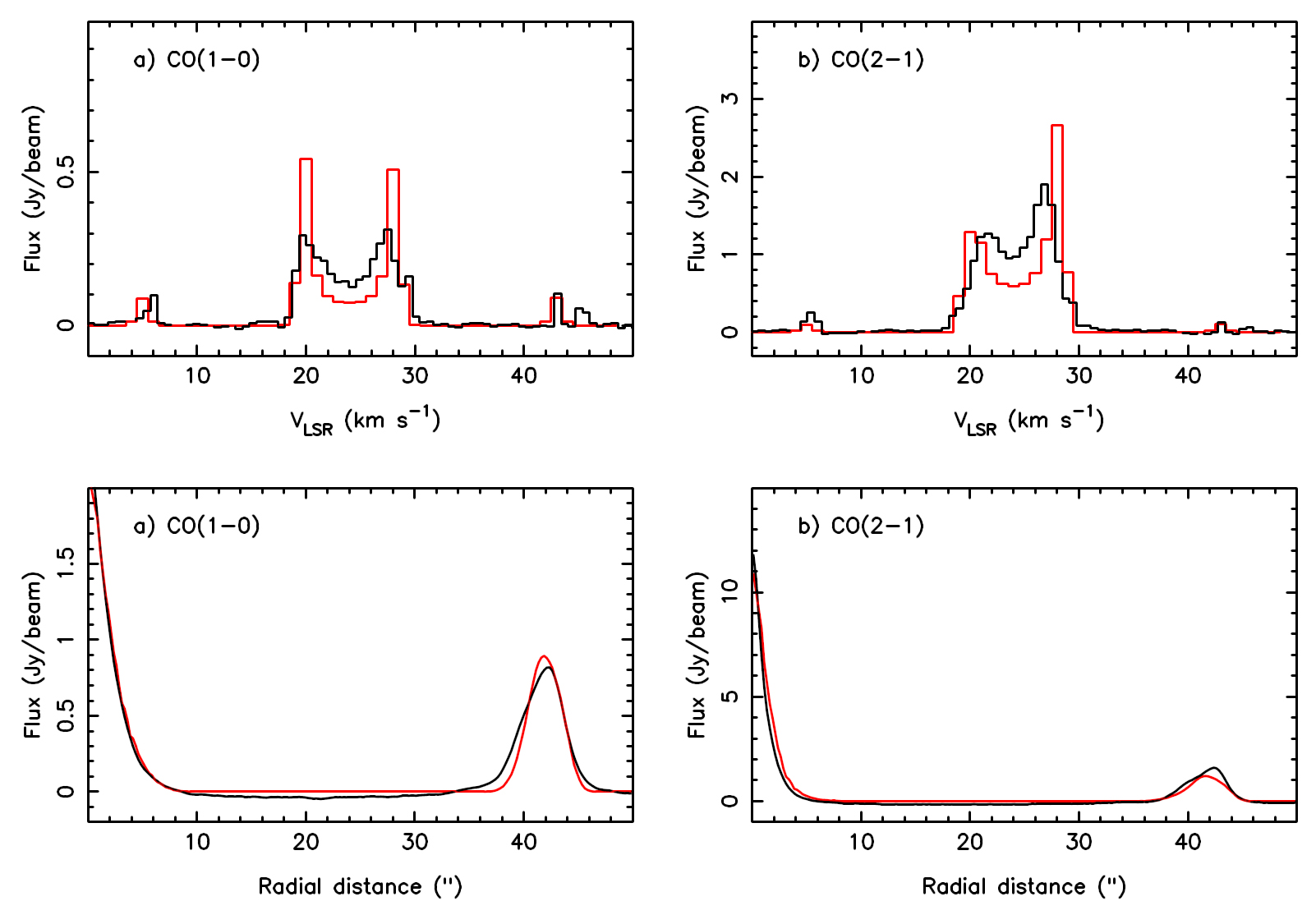}}
  \caption{Upper panels: ALMA $^{12}$CO(1--0) and $^{12}$CO(2--1) line spectra extracted at the stellar position (1.7\arcsec{} and 1.2\arcsec{} annuli, respectively). The red lines show the global best-fit model as discussed in the text. Lower panels: ALMA radial intensity profiles in the $^{12}$CO(1--0) and $^{12}$CO(2--1) lines over-plotted with our best-fit model.}
  \label{F8_model}
  
  \resizebox{\hsize}{!}{\includegraphics[clip, trim=-1cm 0cm -1cm -1cm, width=1.0\textwidth]{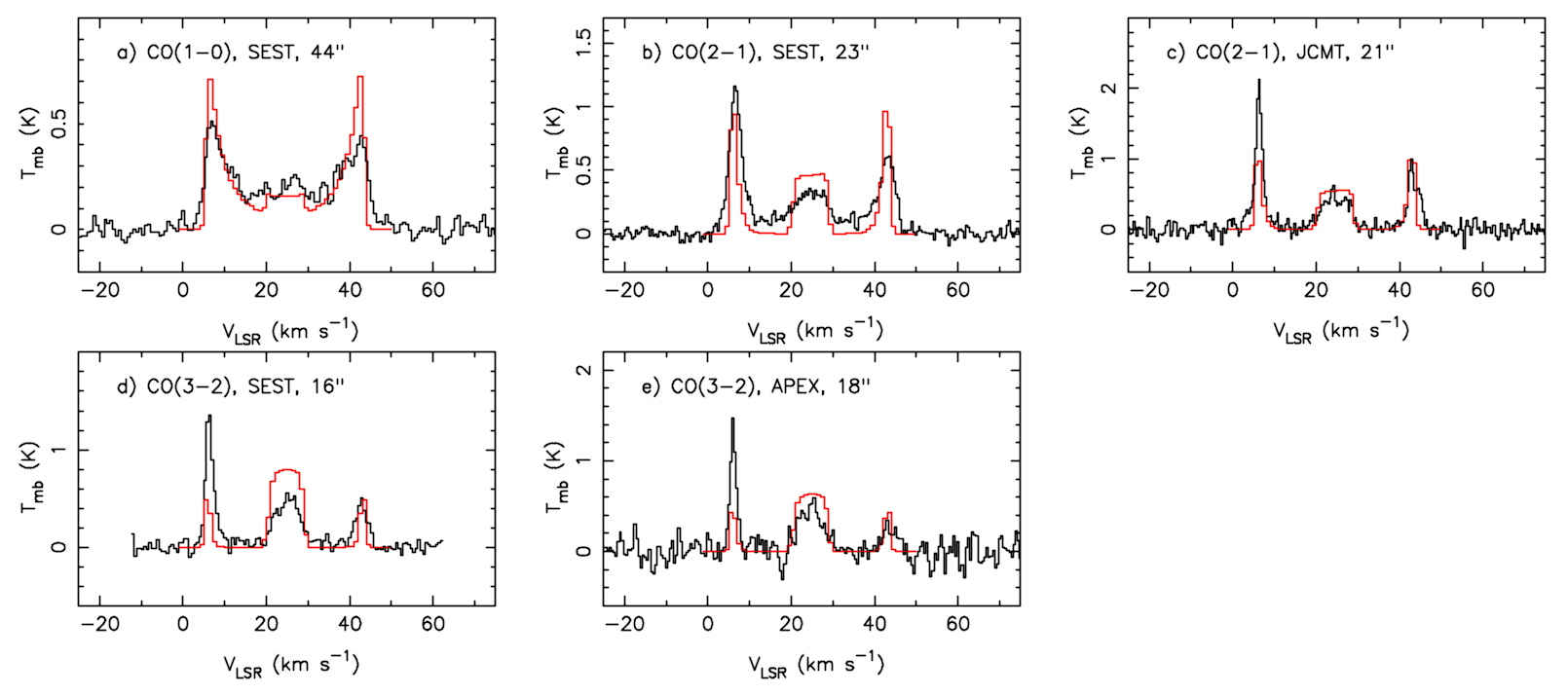}}
  \caption{Single dish observations towards U Ant \citep{Schoeier2005} with transition, telescope, and respective beam sizes indicated. The red lines show the global best-fit model  as discussed in the text.}

  \label{F9_model}
  
\end{figure*}

\citet{Schoeier2005} performed a radiative transfer analysis of single-dish CO data on U Ant in order to derive the present-day mass-loss rate and the mass of the detached shell. We use the same method and code here to model the ALMA data as well as the existing single-dish data in a preliminary analysis (for reasons given below). Following \citet{Schoeier2005}, where the details of the modelling are described, we assume a CO abundance with respect to H$_2$ of 10$^{-3}$, and for the shell a kinetic temperature of 200\,K (not well constrained). A present-day mass-loss rate of 4$\times$10$^{-8}$\,$M_\odot$\,yr$^{-1}$ (twice that of \citet{Schoeier2005}) and a shell mass of 2$\times$10$^{-3}$\,M$_\odot$ (consistent with \citet{Schoeier2005}) result in a reasonably good fit to all the data, including the ALMA radial brightness distributions of the $J$\,=\,\mbox{1--0} and \mbox{2--1} lines, in Figs.~\ref{F8_model} and \ref{F9_model}. 

The good fit to the radial brightness distributions of the ALMA data (lower panels of Fig.~\ref{F8_model})  indicates that the \citet{Mamon1988} results for the photodissociation of CO that are used in the model give a good estimate of the size of the CO envelope produced by the present-day wind for the adopted mass-loss rate. The line intensity contrast between the red- and blue-shifted peaks of the shell emission cannot be reproduced by the model. Its origin probably lies in the large-scale distribution of CO gas in the detached shell, although the very similar appearance of the CO detached shell emission also towards R~Scl and S~Sct may point to another explanation \citep{Olofsson1996}.
    
Taken at face value, the results suggest that the  mass-loss rate has declined considerably since the creation of the shell until the present day. It is presently impossible to say whether this is similar or different to the evolution of the mass-loss rate around R~Scl. In the case of R Scl, the binary-induced spiral made it possible to detect the extended component of the post-pulse wind (i.e. right after the pulse) that may just disappear below the detection threshold without the binary interaction. To have a definite say on this, as well as to improve the modelling of the observed data, requires a much more detailed analysis, for example, a substantial amount of emission from material between the star and the detached shell may be missing due to photodissociation of CO; a consideration of this is planned for a forthcoming paper. 

The mass-loss rate during the shell formation is more difficult to estimate since the shell structure has most likely been affected by hydrodynamical effects during the expansion, which may affect the width of the shell as well as its expansion velocity. Noting these problems and assuming a connection between shell formation and thermal pulses \citep{olofsson1990}, we adopt the timescale of the high-luminosity phase during a thermal pulse, a few\,$\times$\,10$^2$\,yr \citep{karakaslattanzio2007}, as the timescale of the shell formation. In this way, we estimate an average mass-loss rate during the ejection of the shell of the order of $\approx$\,10$^{-5}$\,$M_\odot$\,yr$^{-1}$.

\section{Conclusion}

Spatial and spectral high resolution molecular gas observations from ALMA help us to better understand both the mass-loss morphology and the time evolution of mass loss in the detached shell around the carbon star U~Ant. The detached shell is remarkably thin and overall spherically symmetric. Its size and expansion velocity indicate a short high mass-loss period about 2700 years ago lasting only a few hundred years. When looking into the details of the shell, a filamentary substructure becomes visible that is also accompanied by velocity substructures. The barely resolved filaments are confined to a detached shell width of a differential expansion age of only 150--300\,yr. These timescales fall in the same range as that of high mass loss during a thermal pulse of $\approx$\,200\,yr \citep{Vassiliadis1993} further supporting a short high mass-loss ejection period. The high mass-loss period of U Ant can be characterized by an expansion velocity of the order of 19\,km/s and a mass-loss rate of up to $\approx$\,10$^{-5}$\,$M_\odot$\,yr$^{-1}$. 

These simple conclusions neglect the maybe dominating effects of instabilities in the interaction zone of the fast, dense wind from the thermal pulse that swept up a slower, less dense wind lost prior to the helium-shell flash. The latter may have had similar properties to the low-velocity, present-day wind we observe close to the star. The presence of instabilities in the detached shell is supported by the morphological similarity of the filaments with comparable structures seen in interaction zones of stellar winds with the ISM. Only full hydrodynamical models will be able to clarify the filaments origin. 

The present-day mass loss with a gas expansion velocity of $\approx$\,4.5\,km/s and a mass-loss rate of 4$\times$10$^{-8}$\,$M_\odot$\,yr$^{-1}$ lies at the low end of AGB mass-loss characteristics and seems to have been of that order already for the last 1600\,yrs. Both the expansion velocity difference, and the difference in mass-loss rate between the present-day and the shell ejection epoch are in general agreement with the corresponding evolutionary changes predicted by \citet{Vassiliadis1993} for the interpulse period.

When comparing U Ant to other detached shell objects that have spatially high resolution observations of their circumstellar CO emission, TT Cyg and R Scl, the emission towards U Ant and TT Cyg is consistent with a rapid decline to low gas expansion velocities and mass-loss rates following the thermal pulse. The observations of R Scl indicate a more gradual evolution of the mass- loss characteristics in this source. However, it is only because of the binary-induced spiral shape that this slow decline is made visible in the case of R Scl. Without the spiral shape, the emission around R Scl created by the slowly declining mass-loss rate and expansion velocity would likely look very similar to that of U Ant and TT Cyg.

\medskip

\begin{acknowledgements}

F.K. and M.B. acknowledge funding by the Austrian Science Fund FWF under project number P23586. M.B.further acknowledges funding
through the uni:docs fellowship of the University of Vienna. H.O. acknowledges financial support from the Swedish Research Council. C.P. is supported by the Belgian Fund for Scientific Research F.R.S.- FNRS.

\end{acknowledgements}

\bibliographystyle{aa}
\bibliography{biblio}

\begin{appendix}

\section{Additional figures}

\begin{figure}[h]
   \centering
  \resizebox{\hsize}{!}{\includegraphics[clip, trim=0.1cm 0.1cm 0.1cm 0.1cm, width=1.00\textwidth]{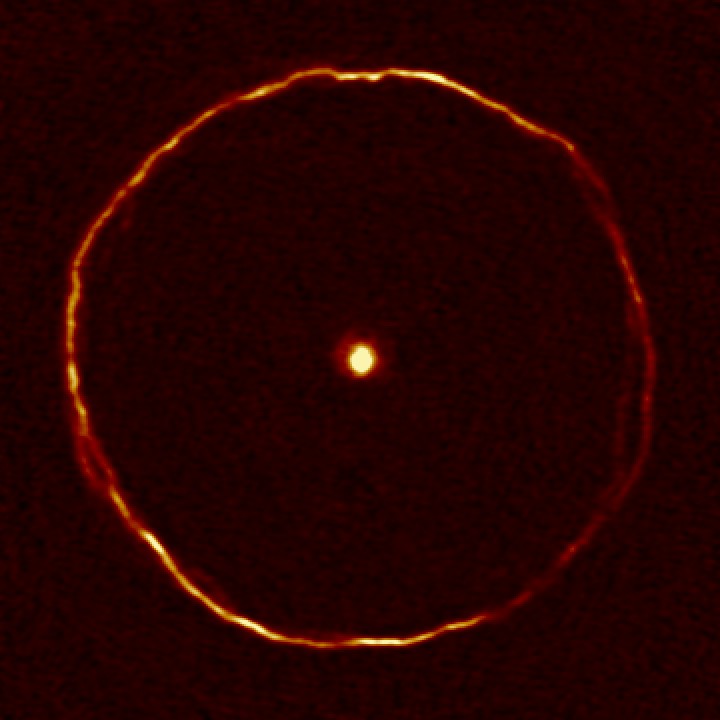}}
   \caption{Single velocity step of the full $^{12}$CO(2--1) animated velocity scan at ({\bf Online movie}).   The velocity resolution and step size is  0.65\,km/s.}
              \label{F6_B6_movie}
    \end{figure}

A full resolution animated velocity scan of the $^{12}$CO(2--1) maps (compare Fig.~\ref{F2_B6_overview}) is shown in  Fig.~\ref{F6_B6_movie}.  It repeatedly scans through all velocity channels at steps of 0.65\,km/s in order to visualize the 3D structure of the detached shell. To more clearly show simultaneously the filamentary structures and the morphology of the region near the star, the movie uses nonlinear colour scales. Figure~\ref{F2_B6_overview} shows the linear flux scales. 
    
\begin{verbatim}
https://dl.dropboxusercontent.com/u/23832864/U_Ant_b6_specloop.mp4
\end{verbatim}

\end{appendix}

\end{document}